\documentclass[%
 reprint,
 amsmath,amssymb,
 aps,
]{revtex4-1}

\usepackage{graphicx}
\usepackage{dcolumn}
\usepackage{bm}
\usepackage{natbib}
\usepackage[table]{xcolor}
\usepackage{tabularx}
\usepackage{array}
\usepackage{colortbl}

\begin{document}

\preprint{APS/123-QED}

\title{Raman Gray Molasses Cooling of Cesium Atoms on the $D_2$ Line}

\author{Ya-Fen Hsiao$^{1,2}$}
\author{Yu-Ju Lin$^1$}
\author{Ying-Cheng Chen$^1$}%
\email{chenyc@pub.iams.sinica.edu.tw }

\affiliation{%
$^1$Institute of Atomic and Molecular Sciences, Academia Sinica, Taipei 10617, Taiwan\\
$^2$Molecular Science Technology, Taiwan International Graduate Program, Academia Sinica and National Central University, Taipei 10617, Taiwan.}

\date{\today}

\begin{abstract}
We present a systematic study on the Raman gray molasses cooling (RGMC) of cesium atoms on the $D_2$ transition. Due to the large splitting in the excited hyperfine transitions of cesium $D_2$ line, it is relatively simple to implement the RGMC with suitable frequency control on the trapping and repumping lasers of the magneto-optical trap (MOT). We have achieved an atom temperature of 1.7$\pm$0.2 $\mu$K after the RGMC with $3.2\times 10^8$ atoms. The phase space density is $1.43\times 10^{-4}$. Compared to the condition with a bare MOT or the MOT with a standard polarization gradient cooling, the phase space density increases by a factor of more than $10^3$ or $10$, respectively.    
\end{abstract}

\pacs{Valid PACS appear here}
\maketitle


\section{Introduction}
Laser cooling and trapping of atoms in magneto-optical traps and optical molasses~\cite {PhysRevLett.59.2631}, developed around the 1980s, has become a starting point for experiments on quantum optics and quantum many-body physics. The sub-Doppler cooling based on the polarization gradients and optical pumping among Zeeman sublevels in optical molasses with a red-detuned cooling laser driving the $F\rightarrow F^{'}=F+1$ cycling transition leads to an atom temperature below the Doppler temperature limit\cite{Dalibard:89,Weiss:89,Lett:89}. Around the middle 1990s, a variant sub-Doppler cooling mechanism was proposed~\cite {GMCth1} and demonstrated~\cite{GMC1,GMC2,GMC3} in which a blue-detuned cooling laser driving the 
$F\rightarrow F^{'}=F$ or $F\rightarrow F^{'}=F-1$ open transition is used. Because atoms can be optically pumped to Zeeman dark states with a significantly reduced fluorescence rate once they are cold, this cooling method is also called "gray molasses cooling" (GMC) while the cooling with a cycling transition is called "bright molasses cooling"(BMC). The GMC allows one to achieve an even colder temperature and a higher atom density compared to the BMC. With either of the sub-Doppler cooling methods, it is advantageous to increase the phase space density of atoms when loading them to either magnetic traps or optical dipole traps for evaporative cooling to quantum degeneracy.     

It is usually considered that sub-Doppler cooling could be ineffective for the $D_2$ transition of lithium and potassium, especially their bosonic isotopes, due to the narrow excited-state structure~\cite{PhysRevA.56.3040}. However, more careful studies show that one can achieve the sub-Doppler temperature by sophisticated dynamic control of the intensity and detuning of the cooling and repumping lasers~\cite{PhysRevA.84.043432,JPB.44.115307}. Recently, the success of blue-detuned, $\Lambda$-enhanced sub-Doppler cooling of $^{40}$K~\cite{40KA0}, $^{39}$K~\cite{K39A0} and $^{7}$Li~\cite{Li7} has attracted renewed interest in laser cooling based on dark states. Under the Raman (two-photon) resonance condition for the two lasers driving the $F\rightarrow F^{'}=F$ and $F-1\rightarrow F^{'}=F$ open transitions, atoms are optically pumped to the dark states formed by coherent superposition of the two ground-state hyperfine manifolds once they are cold. We refer this method as Raman gray molasses cooling (RGMC). The RGMC has been demonstrated in many other atomic species and isotopes, such as $^{6}$Li~\cite{Li6,Li6K40},$^{23}$Na~\cite{Na23,23NaD2}, $^{39}$K~\cite{K39A}, $^{40}$K~\cite{Li6K40,K40B}, $^{41}$K~\cite{K41},  and $^{87}$Rb~\cite{Rb87}. In most of the works with Li, K and Na, the RGMC were implemented with the $D_1$ or the $D_2$ transition which have well-separated hyperfine spacing. Thanks to the large hyperfine splitting in the $D_2$ transition of $^{87}$Rb and $^{133}$Cs, the RGMC could be easily implemented based on the trapping and repumping lasers with suitable frequency control. Although there are other cooling schemes, such as degenerate Raman sideband cooling~\cite{DRSC} that allows one to cool atoms to sub-$\mu$K temperature, they require more lasers and a relatively complicated setup. RGMC provides a simple and effective way to increase the phase space density of the atoms, which facilitates the loading into optical dipole traps for further cooling to quantum degeneracies~\cite{Li6,K41}. In this paper, we report the RGMC of $^{133}$Cs to 1.7$\pm$0.2 $\mu$K with a capture efficiency of $>$80 \% with an initial atom number of 3.7$\times 10^8$. This brings a closure of the demonstration of RGMC for stable alkali species. Table I gives a summary of the alkali species and isotopes that RGMC have been implemented.   

\begin{table*}[tbp]
\centering
\extrarowheight=5pt 
\setlength{\arrayrulewidth}{0.1mm}
\setlength{\tabcolsep}{9pt}
\renewcommand{\arraystretch}{1.3}

\resizebox{0.93\textwidth}{!}{%
\begin{tabular}{|c|c|c|c|c|c|c|c|c|c|c|}
\hline
Atomic Species &$^{6}$Li  &$^{7}$Li  &$^{23}$Na  &$^{23}$Na  &$^{39}$K   &$^{40}$K &$^{40}$K  &$^{41}$K  &$^{87}$Rb &$^{133}$Cs\\ \hline 

Min Temp ($\mu$K) & 40/44 & 60 & 9  & 56   & 6/12  & 20/11  & 48  & 42 & 4  & 1.7  \\ \hline
$D_1$/$D_2$ Line  & {$D_1$} & {$D_1$} & {$D_1$} & {$D_2$} & {$D_1$} & {$D_1$} & {$D_2$} & {$D_1$} & {$D_2$} & {$D_2$}  \\ \hline
Reference  
& {\cite{Li6}/\cite{Li6K40}} & {\cite{Li7}}  & {\cite{Na23}} & {\cite{23NaD2}} 
& {\cite{K39A0}/\cite{K39A}} & {\cite{40KA0}/\cite{Li6K40}}  & {\cite{K40B}}   & {\cite{K41}}  & {\cite{Rb87}} & {This work}
\begin{tabular}[c]{@{}c@{}}
\end{tabular} \\ \hline
\end{tabular}%
}
\caption{A summary of alkali species and isotopes that Raman gray molasses cooling have been implemented. Which transition (either $D_1$ or $D_2$) that RGMC was implemented and the minimum achieved temperature are also shown.}
\label{specices}
\end{table*}

The theoretical aspect of RGMC has been studied in~\cite{Li7,K39A,Rb87}. With the one-dimensional model in a $\Lambda$-type three-level system, one can calculate the friction coefficient and the photon scattering rate based on the optical Bloch equation under certain approximations. A narrow dispersive feature in the friction coefficient around the Raman resonance condition appears with the red(blue)-detuned side of the two-photon  detuning favoring the cooling (heating) force in the case of one-photon blue-detuned for both lasers. The photon scattering rate reaches a minimum at the exact Raman resonance and rises sharply in the blue-detuned side but relative slowly in the red-detuned side. The equilibrium temperature is determined by both the friction coefficient and the photon scattering rate. The degree of Raman coherence of the dark state directly affects the minimum photon scattering rate and thus the atom temperature. As knew in the dark state physics, the mutual laser coherence, the magnetic field, and its inhomogeneity ...etc may affect the Raman coherence~\cite{PhysRevLett.120.183602}. These parameters should be well controlled in order to reach a low atom temperature.

We detail our experimental setup in Sec. II and present the results and discussions in Sec. III, followed by a conclusion. 

\begin{figure*}[ht!] 
\centering 
\includegraphics[width=1.65\textwidth,viewport=10 10 2990 1000,clip]{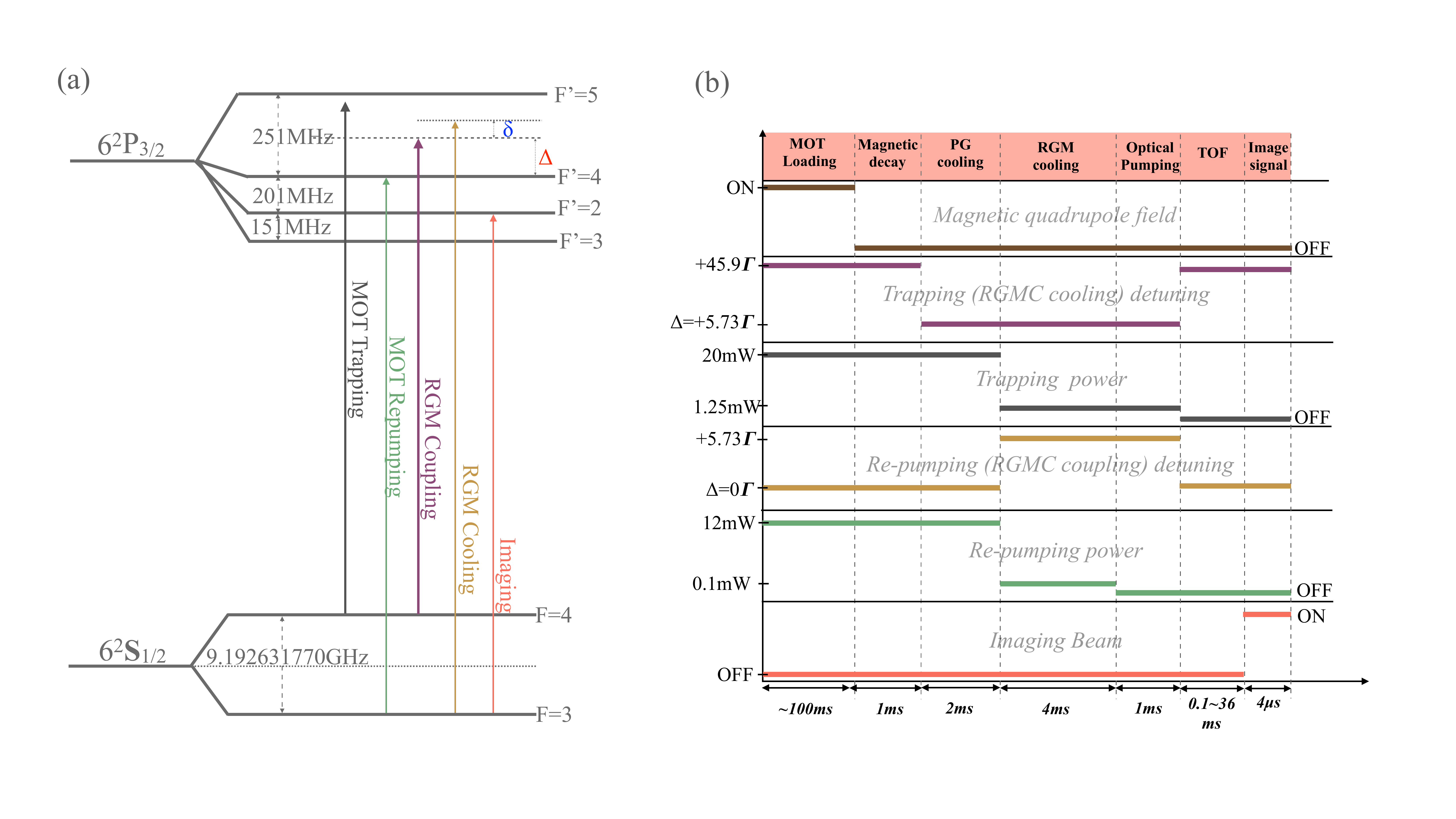} 
\caption{(a) Energy levels of $^{133}$Cs and relevant laser excitations. (b) The timing diagram of the experiment, where $\Delta$ is the detuning w.r.t. $|F=4\rangle\rightarrow|F'=4\rangle$ ($|F=3\rangle\rightarrow|F'=4\rangle$) transition for the RGMC coupling (cooling) light.} 
\label{MOT} 
\end{figure*}

\section{Experimental Setup}
Our cesium magneto-optical trap (MOT) is implemented in a glass cell with six independent trapping and repumping beams. Typically, we have $\sim$20 and $\sim$12 mW per beam for the trapping and repumping light, respectively. The power of the trapping and repumping beams can be tuned by controlling the driving radio-frequency (rf) power of the acousto-optic modulator (AOM) through a voltage-controlled attenuator. The diameters of the trapping and repumping beams are $\sim$23 mm. Two cesium dispensers are put close to the MOT region and operated with a current of $\sim$3 A. Typically, we trap $\sim 4\times10^8$ atoms in the MOT. 

A master laser is locked to the crossover of the $|F=4\rangle\rightarrow|F'=4\rangle$ and $|F=4\rangle\rightarrow|F'=5\rangle$ transitions of cesium $D_2$ line. Part of its light passes through a fiber electro-optic modulator (EOM 1) with the +1 order sideband injection locking an intermediate laser (IL1). Part of the light of IL1 passes through an AOM in a double pass configuration then seeds a tampered amplifier. The output of the amplifier passes one AOM and then coupled into a 2 by 6 fiber beam splitter (OZ Optics NEW FUSED-26-850-5/125-16.7-3S-3-2-PM-SF) to act as the repumping light. The frequency of the repumping light is on the resonance of $|F=3\rangle\rightarrow|F'=4\rangle$ transition under the normal MOT operation.  During the RGMC stage, the frequency of the repumping light jumps to the blue side of the $|F=3\rangle\rightarrow|F'=4\rangle$ transition and acts as the cooling laser for RGMC by controlling the driving frequency of the double-passed AOM. We keep the driving frequency of the EOM 1 fixed although it allows a larger frequency tuning range. This is because that part of the light from EOM 1 is also used for another MOT system in the laboratory and both systems are under operation simultaneously. Another part of the IL1 passes through another fiber-EOM (EOM 2) with its -1 order sideband injection locking one intermediate laser (IL2). The light of IL2 seeds another tampered amplifier. The output of this amplifier passes one AOM and then couples into the 2 by 6 fiber splitter to act as the trapping light. The frequency of the trapping light can be switched to the blue side of the $|F=4\rangle\rightarrow|F'=4\rangle$, whose detuning w.r.t. this transition is denoted as $\Delta$, by changing the driving frequency of fiber EOM 2. It acts as the coupling light for RGMC. The two-photon detuning for the RGMC cooling relative to the coupling light is denoted as $\delta$, as shown in Fig.2(a). 

Three pairs of coils are used to compensate the stray magnetic fields. We use the microwave spectroscopy to diagnose the magnitude of the magnetic field and calibrate the magnetic field per current for each pair of compensation coils. Based on these, it is relatively easy to nullify the stray magnetic field to a few mG level, which is crucial to reach a low atom temperature by RGMC. This is not surprising since minimizing the stray magnetic field effectively reduces the ground-state decoherence rate of the dark states and thus the fluorescence and heating rate.  

Our timing diagram is shown in Fig. 2(b). The experiment runs at a repetition rate of 7.5 Hz. At the end of MOT loading period, the current for the MOT quadrupole magnetic field is turned off within 200 $\mu$s. One ms after turning off the quadrupole magnetic field, the frequency of the trapping laser jumps to the detuning for RGMC, which is typically $\Delta\approx 4.73 \Gamma$ unless specified. For a duration of 2 ms, we take advantage of the standard bright molasses cooling to pre-cool the atoms to $\sim$9 $\mu$K before performing the RGMC. We also check the performance of RGMC directly from MOT without BMC in which the atom temperature is around 120 $\mu$K. We find that the final atom temperatures after RGMC with and without the pre-cooling are almost the same. We choose the timing with the 2-ms pre-cooling in all data taking. Next, the repumping frequency is switched to the desired value for RGMC, remaining a constant for the following 4 ms. The intensities of both trapping and repumping beams are also reduced to varying values for studying their dependence on the performance of RGMC. The trapping and repumping light act as the role of coupling and cooling light during RGMC stage. We denote the intensities of RGMC coupling and cooling beam as $I_{coup}$ and $I_{cool}$, respectively. In the following 1 ms, the RGMC cooling light is turned off and the population is optically pumped by RGMC coupling light to the $F=3$ hyperfine ground state, which is the desired state for future experiment related to electromagnetically induced transparency. We then turn off all lasers for a certain flight time and fire the imaging beam which drives the $|F=3\rangle\rightarrow|F'=2\rangle$ transition. 

From the absorption imaging we determine the two-dimensional profile (referred as $x$ and $z$, where $z$ is the gravity direction) of the column density. To allow a quicker and more reliable fitting, we sum the column density along $x$ and $z$-axis, respectively. The results are fitted to one-dimensional Gaussian function to get the $e^{-1}$ width in each axis ($\sigma_i$, i=$x,z$). Based on the fit width and amplitude, we obtain the atom number using an absorption cross-section of $\frac{5}{21}\frac{3\lambda^2}{2\pi}$ assuming a uniform population in Zeeman sublevels and with a linearly-polarized image beam, where $\lambda=852.35$~nm is the wavelength of the $D_2$ transition. The atom temperature can be determined by fitting the $e^{-1}$ width of atomic clouds versus different flight times with the formula,
\begin{equation}
\sigma_i(t_{TOF})=\sqrt{\sigma_{i,0}^2+\frac{2k_BT_i}{m}t_{TOF}^2}.
\label{tof}
\end{equation}
The maximum flight time can be up to 36 ms for a temperature of $\sim$2 $\mu$K. The position shift of the atoms due to gravitational free-fall motion is used to measure the magnification ratio of the imaging system, allowing an accurate determination of the atom temperature. Our atomic clouds are in elliptical shape with the width $\sigma_x$ 2-3 times larger than $\sigma_z$. The fitting result of $T_x$ is not as reliable as $T_z$ since it needs even longer flight times to allow a sufficient expansion in cloud size. But the images at longer flight times are limited by the size of the CCD chip. The atom temperatures shown in all figures are $T_z$.    
\section{Results and Discussions}
We have performed a systematic study on the atom number and temperature dependence on various parameters during the Raman gray molasses cooling.     
\subsection{\label{sec:citeref}Cooling time dependence}
\begin{figure}[h!] 
\centering 
\includegraphics[width=1.95\textwidth,viewport=50 10 2890 590,clip]{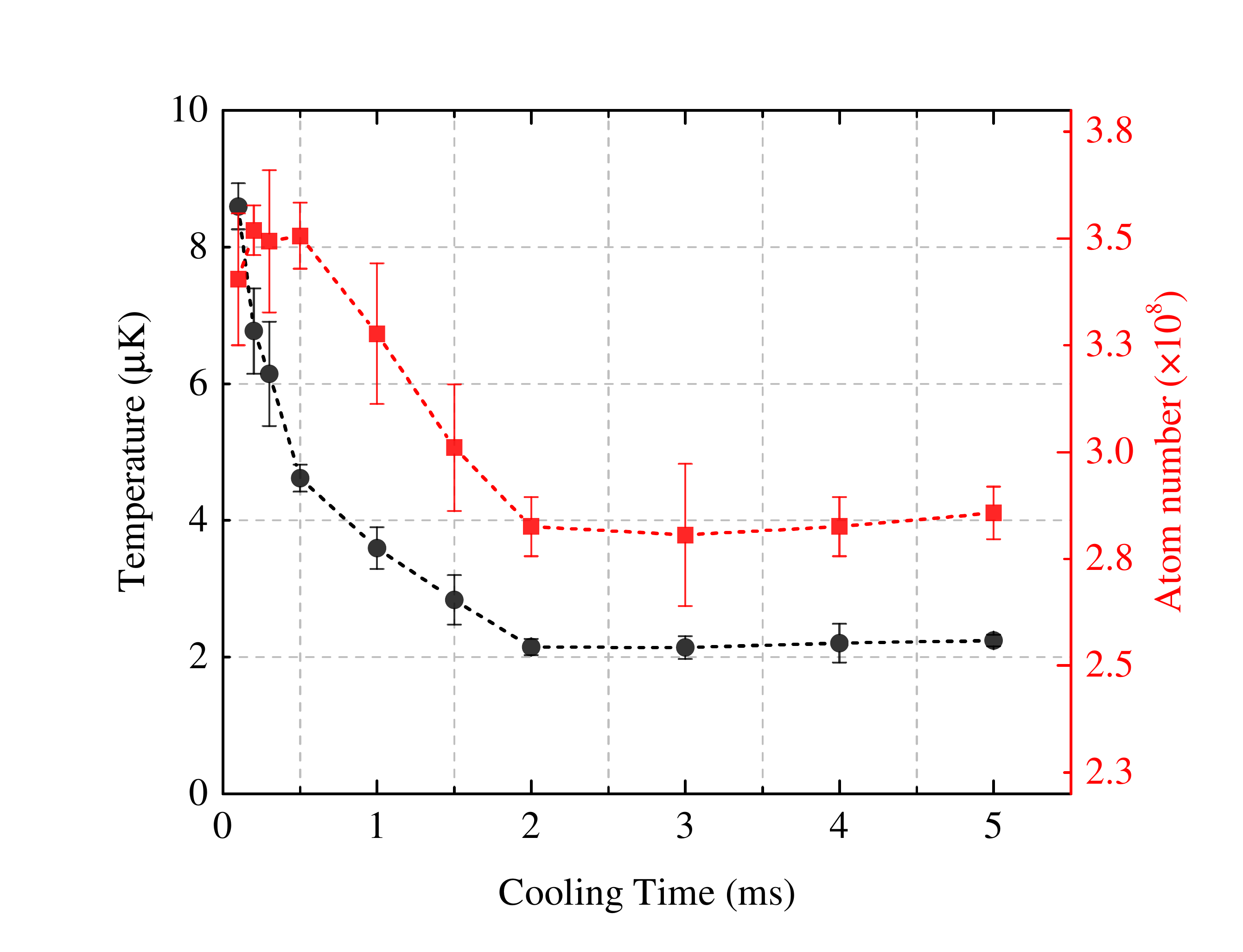} 
\caption{Atom number (square) and temperature (circle) versus the Raman gray molasses cooling time. $\delta$=0 and $\Delta$=5.73 $\Gamma$. $I_{\rm coupling}$= 0.273$I_{\rm sat}$ and $I_{\rm cool}/I_{\rm coup}=0.08$. }
\label{Cooling_time} 
\end{figure}

We first study the atom number and temperature dependence versus the RGMC time at a zero two-photon detuning $\delta$ and a one-photon detuning $\Delta$ of 5.73$\Gamma$, as shown in Fig.\ref{Cooling_time}. During RGMC, the one-beam coupling and cooling intensities ($I_{\rm coup}$ and $I_{\rm cool}$) are 0.273 and 0.022 $I_{\rm sat}$, respectively, where $I_{\rm sat}=1.10$~mW/cm$^2$ is the saturation intensity of cesium $D_2$ transition. Prior the RGMC, the atoms have been cooled to 8.4 $\mu$K by the bright molasses cooling for 2 ms. With a RGMC time of $>\sim$2 ms, both atom temperature and number reach a steady-state value. We observed a $\sim$ 28\% reduction in atom number and a fourfold reduction in temperature at the steady state. Because we decrease the laser power suddenly to a lower value, the cooling force also decreases significantly and is not sufficient to capture atoms with larger velocities. This causes some atom losses during the molasses stage. This atom loss could be avoided if one gradually decreases the laser power~\cite{Na23}, although we did not implement it in this work. In all of the following studies, we choose an RGMC duration of 4 ms.   

\subsection{Magnetic field dependence}
\begin{figure}[h!] 
\centering 
\includegraphics[width=1.95\textwidth,viewport=50 10 2890 590,clip]{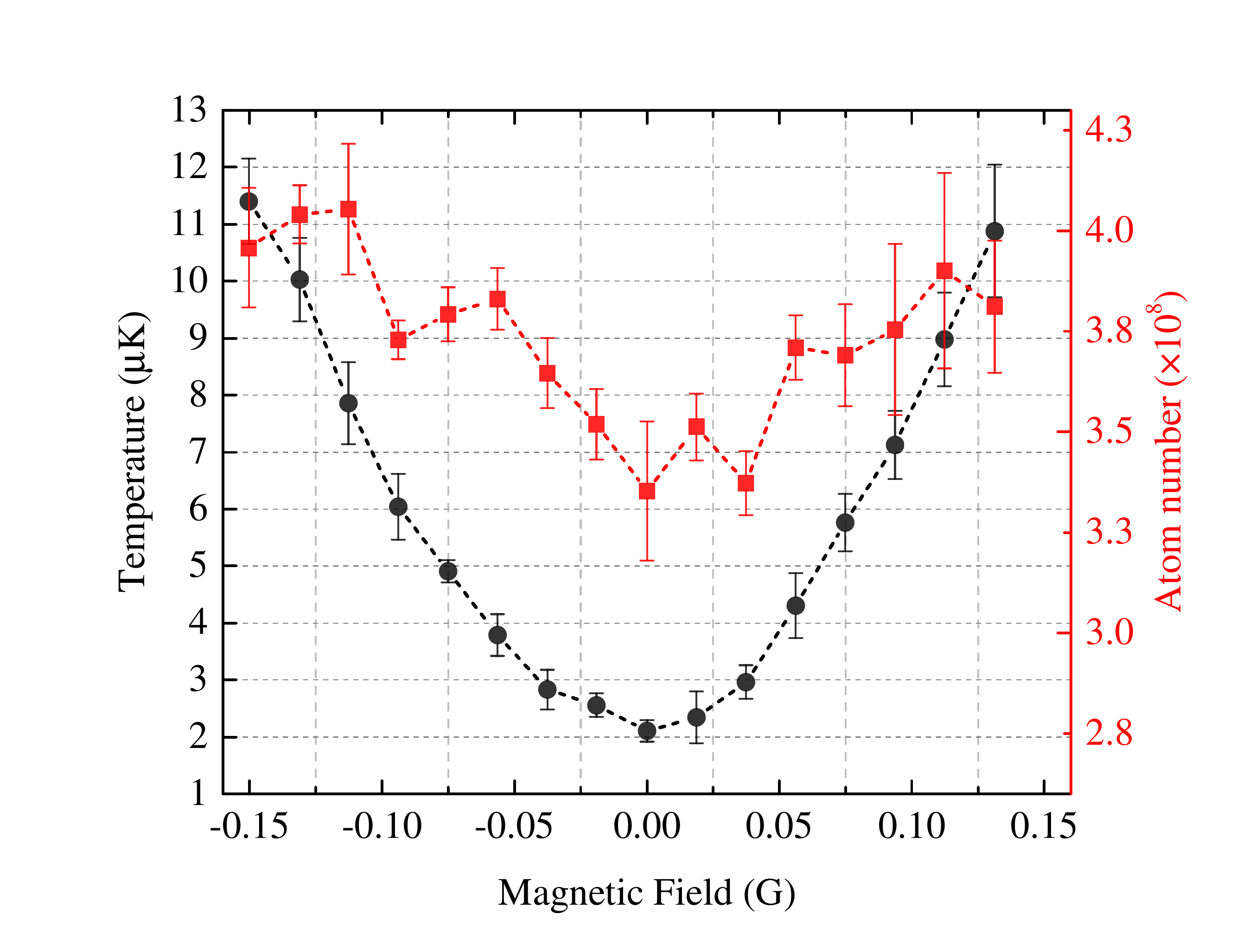} 
\caption{Atom number (square) and temperature (circle) after 4-ms RGMC versus the magnetic field. $\delta$=0 and $\Delta$=5.73$\Gamma$. $I_{coup}$= 0.273$I_{sat}$ and $I_{\rm cool}/I_{\rm coup}=0.08$. } 
\label{Bias_B_field} 
\end{figure}

Because RGMC utilizes the dark state, it is expected that the ground-state decoherence rate plays an important role in the cooling performance. During the cooling period, atoms may distribute among different Zeeman sublevels. Minimization of the stray magnetic field effectively reduces the distribution in Zeeman shifts and thus the ground-state decoherence rate. A smaller decoherence rate for the dark state results in a smaller fluorescence (heating) rate and thus a lower atom temperature. Fig. \ref{Bias_B_field} depicts the atom number and temperature versus the magnetic field by controlling the current through one pair of the compensation coils. The atom temperature shows a quadratic dependence on the magnetic field~\cite{40KA0} as $\Delta$T=532$B^2\mu K/G^2$. The stray magnetic field needed to be canceled to less than 14 mG in order to have a negligible increase ($< 5\%$) in temperature. 

Another important factor that affects the decoherence rate of the dark state is the mutual coherence between the RGM coupling and cooling lasers. It has been studied that the laser mutual coherence significantly affect the minimum temperature~\cite{Rb87}. The minimum temperature achieved with the cooling and coupling lasers being injection-locked to the same master laser is much lower than that achieved with the two lasers independently locked~\cite{Rb87}. Since our RGMC cooling and coupling lasers are injection locked to the same master laser with frequency offsets determined by EOMs or AOMs driven by stable signal generators (stabilities $<10^{-10}$), the laser mutual coherence should not be the major limitation in the minimum temperature.  

\subsection{Intensity dependence}
\begin{figure}[h!] 
\centering 
\includegraphics[width=1.55\textwidth,viewport=40 10 2290 590,clip]{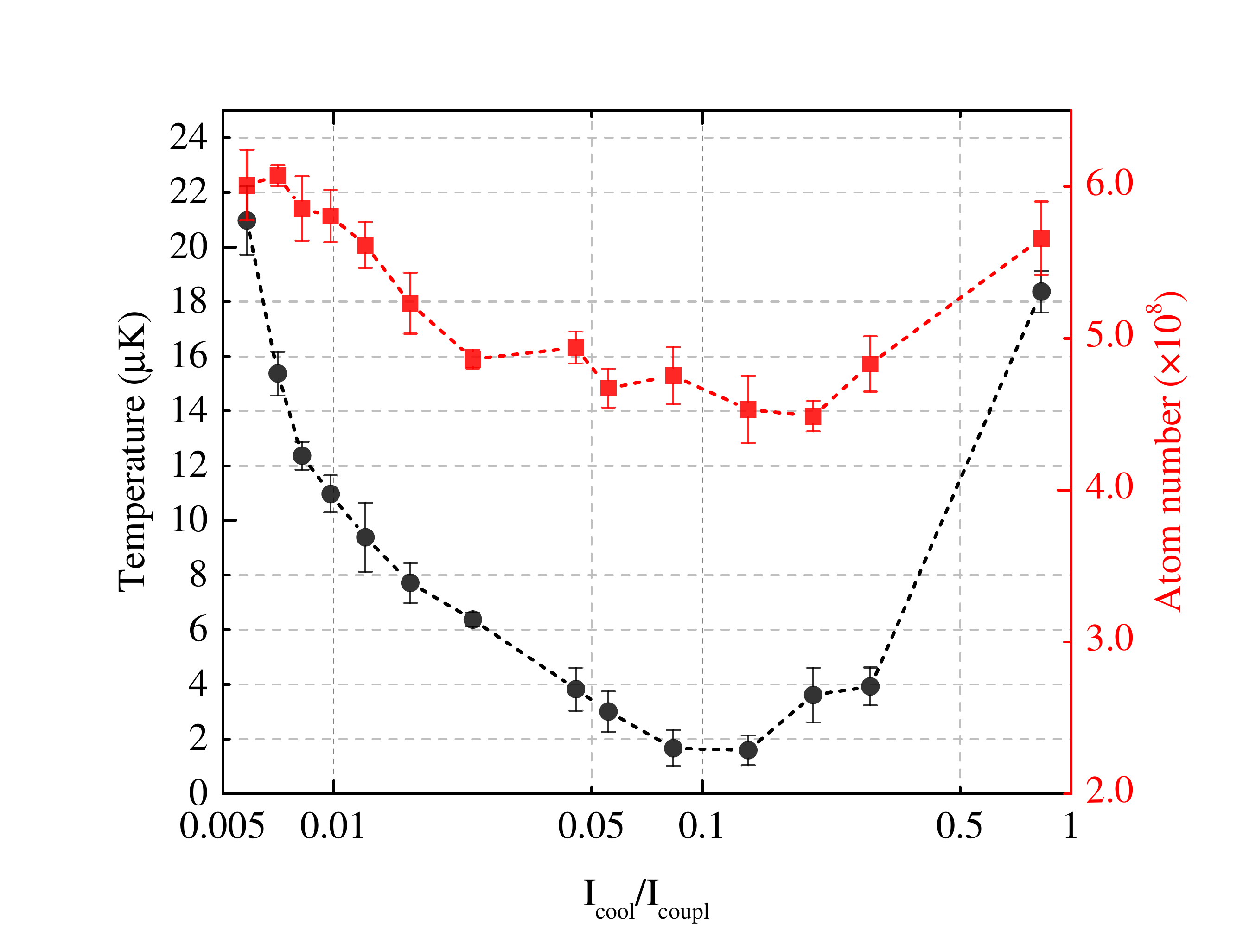} 
\caption{Atom number (square) and temperature (circle) after 4-ms RGMC versus the intensity ratio of RGMC  cooling to coupling beam. $\delta$=0 and $\Delta$= 5.73$\Gamma$. $I_{cool}$=0.022$I_{sat}$.} 
\label{Vary_Power_Ratio} 
\end{figure}

At a one-photon detuning $\Delta$ of 5.73$\Gamma$ and a zero two-photon detuning $\delta$, we vary the intensity of RGMC coupling beam ($I_{\rm coup}$) while keeping the RGMC cooling beam intensity ($I_{\rm cool}$) fixed at 0.023$I_{\rm sat}$ (per beam) and measure the atom number and temperature dependence on the intensity ratio ($I_{\rm cool}/I_{\rm coup}$). The results are shown in Fig. \ref{Vary_Power_Ratio}. We find an optimal intensity ratio ($I_{\rm cool}/I_{\rm coup}$) around 0.1 where atom temperature reaches a minimum.     

\begin{figure}[h!] 
\centering 
\includegraphics[width=1.95\textwidth,viewport=40 10 2890 590,clip]{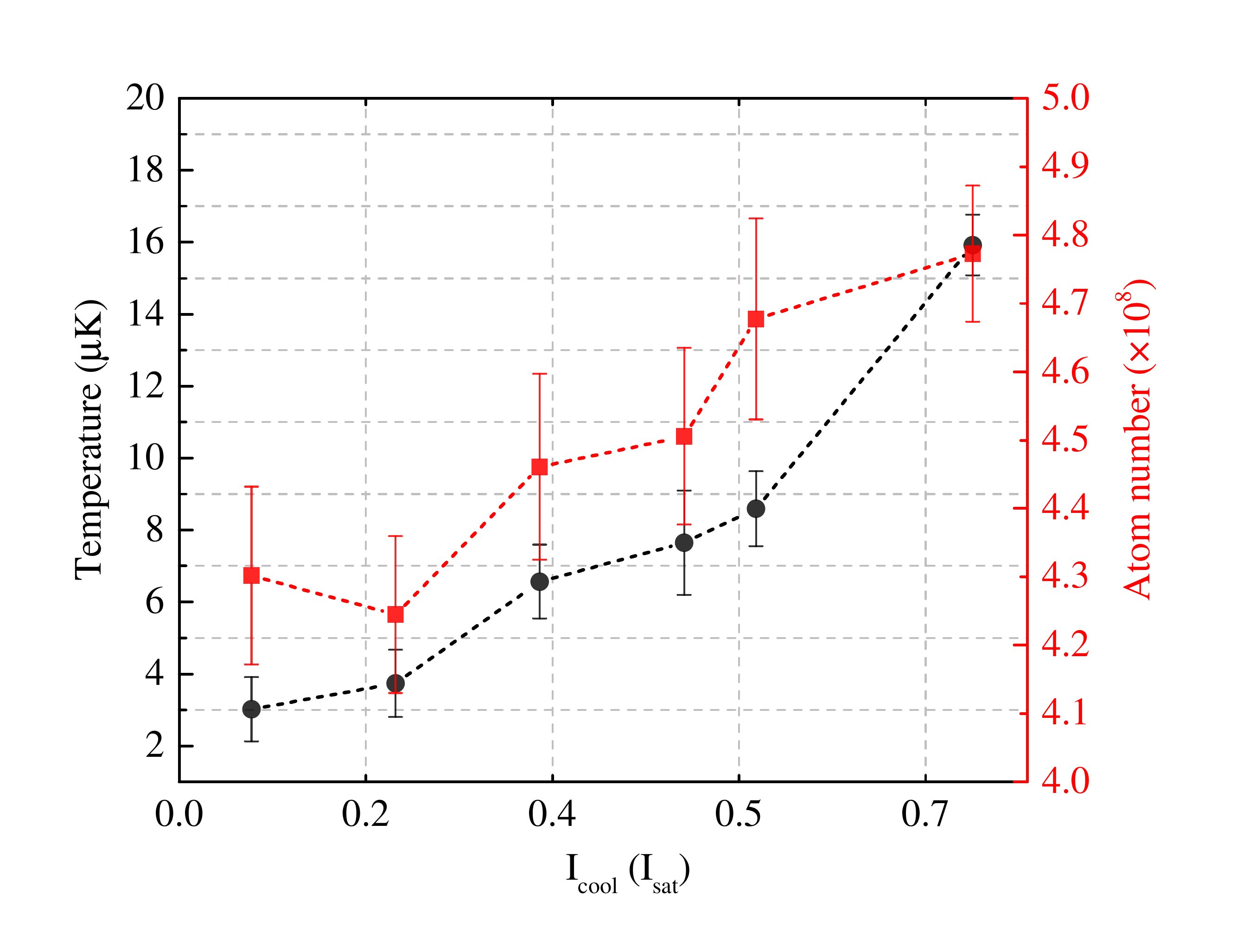} 
\caption{Atom number (square) and temperature (circle) after 4-ms RGMC versus the intensity of RGMC coupling beam. $I_{\rm cool}/I_{\rm coup}=0.08$. $\delta$=0 and $\Delta$=5.73$\Gamma$.} 
\label{power_ratio} 
\end{figure}

We then keep the intensity ratio fixed at 0.08 and vary the power of both beams during the RGMC period. The atom number and temperature versus the RGMC cooling beam intensity are shown in Fig.\ref{power_ratio}. It shows that atom temperature is monotonically proportional to laser intensity. The atom number reduces slightly for lower laser intensities. This is expected as atoms with larger velocities may leave the trap due to weaker trapping forces.  

\subsection{Detuning dependence}

\subsubsection{Two-photon detuning}
\begin{figure}[h!] 
\centering 
\includegraphics[width=1.55\textwidth,viewport=40 10 2290 590,clip]{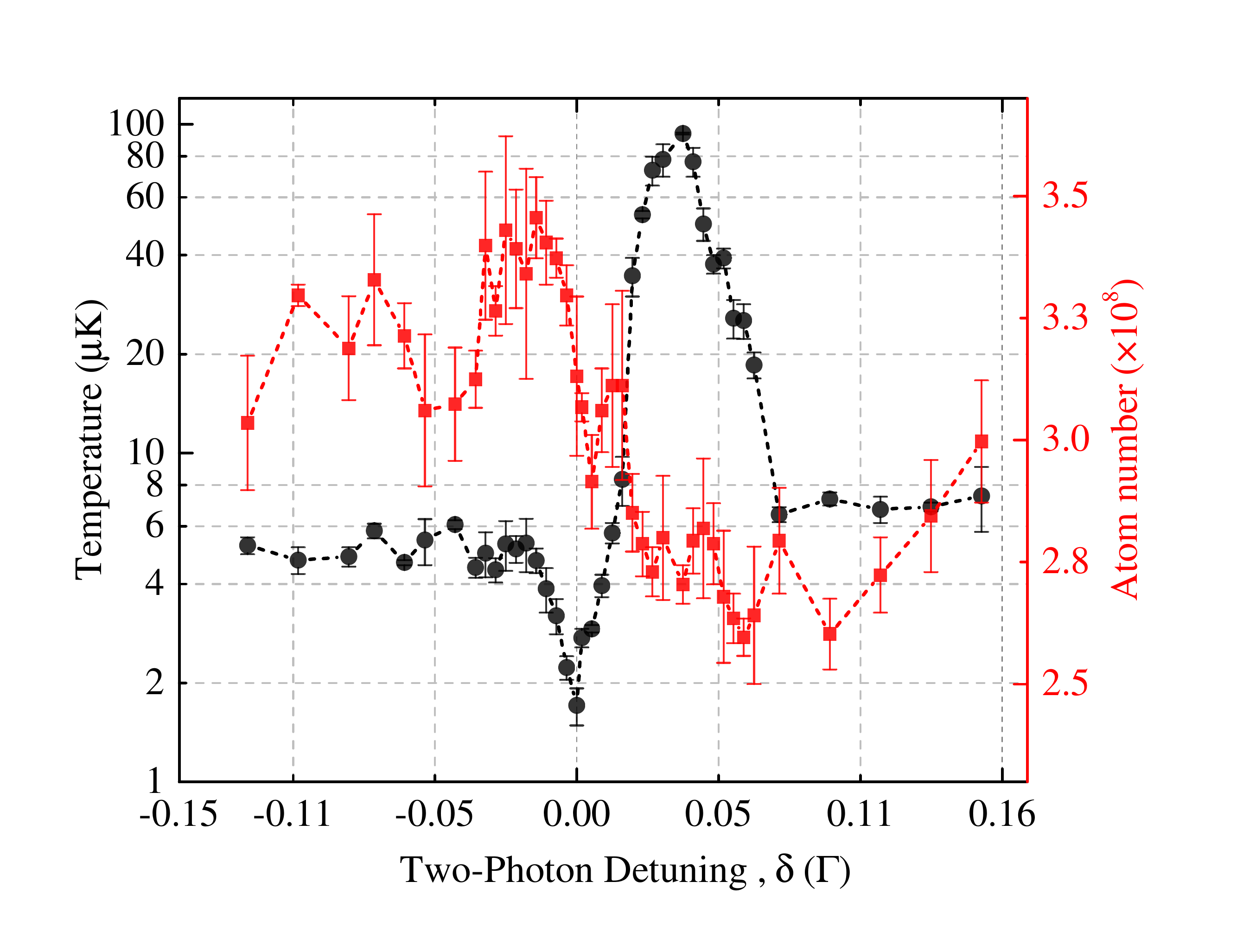} 
\caption{Atom number (square) and temperature (circle) after 4-ms RGMC versus the two-photon detuning $\delta$. $\Delta$=5.73 $\Gamma$. $I_{\rm coup}$= 0.273 $I_{\rm sat}$ and $I_{\rm cool}/I_{\rm coup}=0.08$.} 
\label{Two_photon_detuning} 
\end{figure}

At a $\Delta$ of 5.73$\Gamma$ and an intensity ratio of 0.08, we vary the detuning of the RGMC cooling beams and measure the atom number and temperature versus the two-photon detuning $\delta$, as shown in Fig. \ref{Two_photon_detuning}. The lowest temperature of 1.7$\pm$ 0.2 $\mu$K appears at $\delta$=0, which is a characteristic for Raman gray molasses cooling. At $\delta$=0, the captured atom number is $\sim$ 80\% of that before RGMC, which is $3.7\times 10^8$. For blue detuning ($\delta>$0), the atom temperature rises sharply up to $\sim$100 $\mu$K while the atom number drops slightly. The phase space density at the lowest temperature is $1.43\times10^{-4}$. Compared to the condition before RGMC, it increases by more than a factor of 10. Compared to the bare MOT without pre-cooling which has a temperature of 120 $\mu$K, the phase space density increases by more than a factor of 1000. 

\subsubsection{One-photon detuning}

\begin{figure}[ht]
\centering 
\includegraphics[width=1.95\textwidth,viewport=50 10 2890 590,clip]{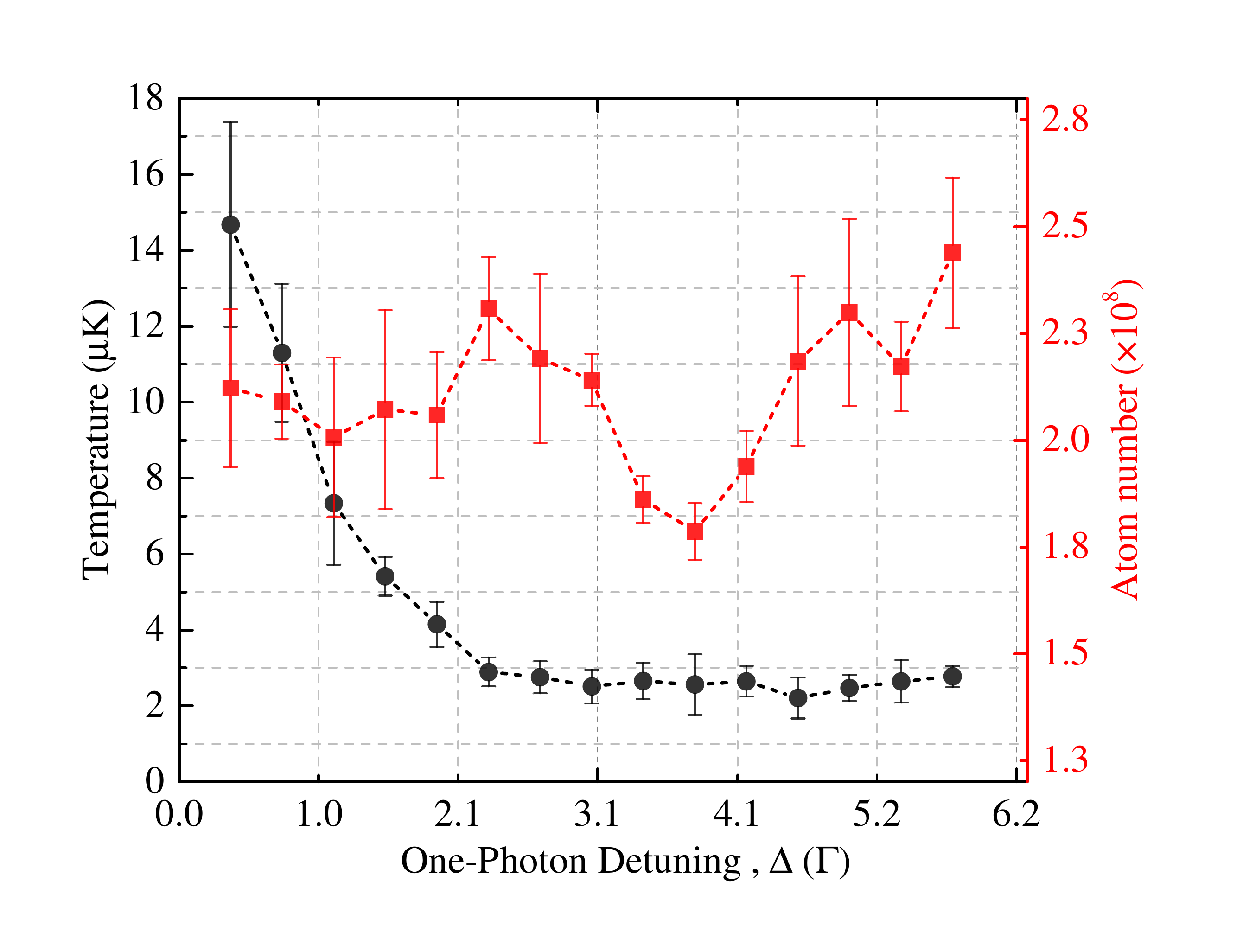} 
\caption{Atom number (square) and temperature (circle) after 4-ms RGMC versus the one-photon detuning $\Delta$. $\delta$=0. $I_{\rm coup}$= 0.273 $I_{\rm sat}$ and $I_{\rm cool}/I_{\rm coup}=0.08$.}
\label{one_photon}
\end{figure}

We then keep the two-photon detuning fixed at zero and measure the atom number and temperature dependence on the one-photon detuning $\Delta$, as shown in Fig. \ref{one_photon}. The temperatures decrease and reach an almost constant value as the $\Delta$ increase. Limited by the frequency tuning range of the double-passed AOM, the maximum $\Delta$ is 5.73 $\Gamma$. Similar behaviors were observed in many other works~\cite{K40B,Rb87,23NaD2}. As $\Delta$ increase further, we expect the temperature will rise up at some point since the RGMC coupling laser may drive the $|F=4\rangle\rightarrow|F'=5\rangle$ cycling transition significantly and degrade Raman coherence of the dark states due to spontaneous decay\cite{23NaD2}. 

\section{Conclusion}
In conclusion, we perform a systematic study on Raman gray molasses cooling of cesium with the $D_2$ line. The lowest atom temperature of 1.7$\pm$0.2 $\mu$K is achieved. The RGMC provides a simple and effective way to increase the phase space density for further magnetic or optical dipole trap loading and evaporative cooling towards quantum degeneracy.

\section{Acknowledgements}
We thank Ming-Shien Chang for having useful discussions. We acknowledge the financial support from Ministry of Science and Technology of Taiwan under Grant No. 103-2112-M-001-010-MY3 and 106-2119-M-001-002. We also thank the supports from National Center for Theoretical Sciences and Center for Quantum Technology of Taiwan. 

\section*{Reference} 
\bibliographystyle{apsrev4-1} 
\bibliography{ref}

\newpage

\end{document}